\documentclass[aps,pre,showpacs,superscriptaddress,preprintnumbers,amsmath,amssymb]{revtex4}
\usepackage{graphicx}
\usepackage{dcolumn}
\usepackage{bm}
\usepackage{float}

\begin{document}
\title{Target shape effects on monoenergetic GeV proton acceleration}
\author{Min Chen} \affiliation{Institut f\"{u}r Theoretische
Physik I, Heinrich-Heine-Universit\"{a}t D\"{u}sseldorf, 40225
D\"{u}sseldorf, Germany}
\author{Tong-Pu Yu}
\affiliation{Institut f\"{u}r Theoretische Physik I,
Heinrich-Heine-Universit\"{a}t D\"{u}sseldorf, 40225 D\"{u}sseldorf,
Germany}\affiliation{Department of Physics, National University of
Defense Technology, Changsha 410073, China}
\author{Alexander Pukhov}
\thanks{To whom any corresponding should be sent: pukhov@tp1.uni-duesseldorf.de}
\affiliation{Institut f\"{u}r Theoretische Physik I,
Heinrich-Heine-Universit\"{a}t D\"{u}sseldorf, 40225 D\"{u}sseldorf,
Germany}
\author{Zheng-Ming Sheng}
\affiliation{Department of Physics, Shanghai Jiao Tong University,
Shanghai 200240, China \\and Beijing National Laboratory of
Condensed Matter Physics, Institute of Physics, Beijing 100080,
China}

\begin{abstract}
When a circularly polarized laser pulse interacts with a foil
target, there are three stages: pre-hole-boring, hole-boring and the
light sail acceleration. We study the electron and ion dynamics in
the first stage and find the minimum foil thickness requirement for
a given laser intensity. Based on this analysis, we propose to use a
shaped foil for ion acceleration, whose thickness varies
transversely to match the laser intensity. Then, the target evolves
into three regions: the acceleration, transparency and deformation
regions. In the acceleration region, the target can be uniformly
accelerated producing a mono-energetic and spatially collimated ion
beam. Detailed numerical simulations are performed to check the
feasibility and robustness of this scheme, such as the influence of
shape factors and surface roughness. A GeV mono-energetic proton
beam is observed in the three dimensional particle-in-cell
simulations when a laser pulse with the focus intensity of
$10^{22}W/cm^2$ is used. The energy conversion efficiency of laser
pulse to accelerated proton beam is more than $23\%$. Synchrotron
radiation and damping effects are also checked in the interaction.
\end{abstract}


\maketitle

\section{INTRODUCTION}
Ion acceleration is one of the most important topics in the ultrashort
ultraintense (USUI) laser-plasma accelerator
field~\cite{TNSA,shock-acce,Esirkepov2004,other-acce}. Because
the laser-accelerated proton and ion beams are highly concentrated,
ultrashort and ultraintense, they have a broad spectrum of
potential applications, such as proton therapy~\cite{therapy}, proton
imaging~\cite{ionimaging}, ion beam ignition of laser fusion
targets~\cite{fusion}, \emph{etc}.
These applications usually require a high degree of beam
monochromaticity as well.

The usual energy spectrum of the ion beams produced due to the
mechanism of target normal sheath acceleration~(TNSA) shows an
exponential decay up to a high energy cutoff. To overcome this, a
few schemes are proposed such as use of plasma
lens~\cite{lens,lenstheory}, double layer
targets~\cite{double-layer} and special target
shaping~\cite{Schwoerer2006,Bulanov2002,Okada2006,Robinson2007}.

It was shown recently  in theoretical
analysis and numerical simulations that when a circularly
polarized~(CP) pulse is used, high energy
mono-energetic ion acceleration can be achieved. The reason is
absence of the oscillatory term in the ponderomotive force of a CP
pulse and thus
suppression of the electron heating~\cite{Macchi2005}. A CP laser
pulse pushes gently the whole electron fluid and generates a moving local
charge separation field that accelerates ions. It looks as if the whole
target is accelerated by the laser radiation
pressure. The mechanism then is similar to the radiation pressure
dominated acceleration~(RPDA) proposed by Esirkepov \textit{et al.}
before~\cite{Esirkepov2004}. However, the threshold for the laser
intensity is dramatically reduced when a CP pulse is used. In the one
dimensional geometry and if the laser pulse is long enough, the target
can be accelerated in multi stages and in every stage the
accelerated ions gain the same energy. The narrow
width of the final energy spectrum is a
result~\cite{Zhang2007-multi}. Certainly, the mechanism is sensitive
to a laser prepulse as it might cause pre-plasma at the target surface
and destroy the accelerating structure.
Thanks to the development of USUI laser pulse and plasma mirror
technology, this can be controlled in today's experiments.

Recently, Klimo \textit{et al.}~\cite{Klimo2008}, Robinson
\textit{et al.}~\cite{Robinson2008} and Yan \textit{et
al.}~\cite{Yan2008} have studied an ultra thin target condition.
They have shown that ions are accelerated continuously and rotate in
the phase space. This can be considered as a special kind of multi
stage acceleration~\cite{Zhang2007}. The second stage begins before
the end of the first one. The width of the spectrum is further
reduced and mono-energetic ion beams are easily observed. Qiao
\textit{et al.}~\cite{Qiao2009} recently studied this kind of
acceleration and separated it into two different processes: hole
boring and light sail. They noticed the importance of the transition
between these two processes to reach the final mono-energetic
spectrum. In the hole boring process, the laser pulse propagates
thorough the bulk of the foil target and compresses it. Afterwards,
when the compressed layer arrives at the rear side of the foil, the
light sail process begins. At this stage, the whole target dynamics
can be well described using a ballistic
equation~\cite{Tripathi2009}. There is a minimum target thickness
for the target to stay opaque so that the process can operate. To
get the minimum target thickness for acceleration, all of the
earlier works use the balance between the laser pressure and the
charge separation field and assume immobile ions during the balance
build-up. This is correct when the laser intensity is relatively
small and the target density is low. However, it overestimates the
minimum target thickness when the laser intensity and target density
increase.

In this paper we study the process of the balance build-up and name
this process as a pre-hole-boring process. We consider the finite
ion mass to calculate the minimum foil thickness for ion
acceleration, which is usually also the optimal thickness. Then, we
suggest to use a shaped target to achieve a uniform acceleration.
For shaped targets, we find that the acceleration structure can be
kept for a longer time compared with usual flat targets. The final
spectrum contains a mono-energetic peak. Later we study the
robustness of our scheme by considering the influence of the shape
parameters and surface roughness on the final spectrum. This affords
a detailed description to our recently published
paper~\cite{Chen2009}. We also make some consideration on the
radiation damping effect during the ion acceleration. Finally, we
give some discussions and a summary.

\section{Pre-hole-boring and optimal target thickness}

\begin{figure}\suppressfloats
\includegraphics[width=9cm]{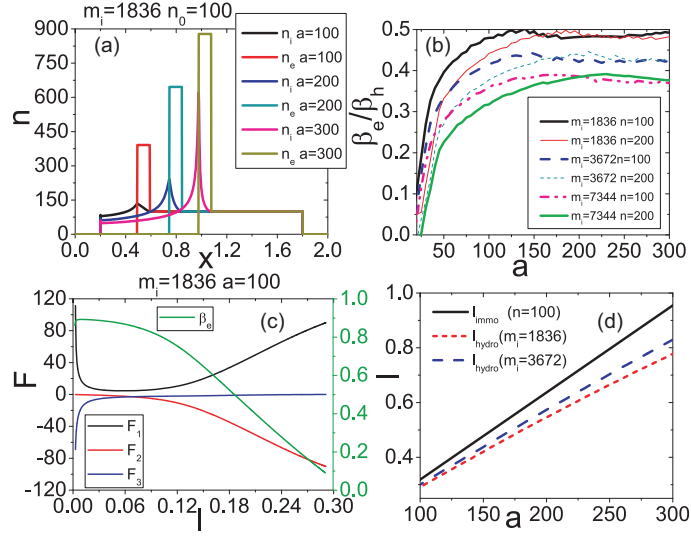}%
\caption{\label{pre-hole-boring} Analytical results for the
pre-hole-boring process. (a) Density distribution of the electrons
and ions at the end of the numerical calculation. (b) The dependance
of final velocity of the CEL at the end of the calculation on the
laser intensity, target density and ion mass. (c) Evolution of CEL
velocity and forces on the CEL along with the CEL
displacement($l_1$). (d) CEL displacements at the end of the
calculation for different ion masses and laser intensities. The
black solid line represents the minimum thickness obtained in an
immobile ion model.}
\end{figure}

Here, we study the electron and ion dynamics during the laser
impinging. When the laser pulse arrives at the front surface of the
target, electrons are accelerated and piled up by the laser
pressure. Later, ions start to move being accelerated by the charge
separation field. We call this stage pre-hole-boring since during
this time ions have not caught up with the compressed electron
layer~(CEL), and the hole boring process did not reach its
stationary stage yet.

For convenience we use normalized variables here. The laser electric
field is normalized as $a=eE/m{\omega}c$, spatial and temporal
coordinates are normalized by the laser wavelength $\lambda$ and
period $2\pi/\omega$, respectively. The particle velocity, mass,
plasma density are normalized by light speed in vacuum $c$, electron
mass $m$, and critical density $n_c=m\omega^2/4{\pi}e^2$,
respectively.

If we consider only the ponderomotive force of the laser pulse
acting on the plasma electrons (the light pressure), the dynamics of
the CEL is governed by the equations:

\begin{eqnarray}\label{CEL}
n_0l_1\frac{d\gamma\beta}{dt}+n_0\gamma\beta\frac{dl_1}{dt} &=&
\frac{a^2_0}{\pi}\frac{1-\beta}{1+\beta}-{\pi}n_0^2l_1(l_1+l_s)\\
\frac{dl_1}{dt} &=& \beta,
\end{eqnarray}

\noindent where $\beta$ is the normalized velocity of the CEL, $\gamma$ is the
relativistic factor, $l_1$ is the displacement of the CEL, $n_0$ is
the initial plasma density, $a_0$ is the laser electric field and
$l_s$ is the skin length for the laser pulse. The second term on the
left side of the first equation comes from the mass increase of the
CEL. The first term on the right side is the contribution of the
laser pressure and the second one is due to the charge separation
field.

We treat ions as a hydrodynamical fluid and solve the
following hydrodynamical equations:

\begin{eqnarray}
\frac{{\partial}n}{{\partial}t}+\frac{{\partial}(n\beta_i)}{{\partial}t}
&=& 0, \\
\frac{{\partial}\beta_i}{{\partial}t}+\beta_i\frac{{\partial}\beta_i}{{\partial}x}
&=& \frac{2\pi}{m_i}E_x, \\
\frac{{\partial}E_x}{{\partial}x} &=& 4{\pi}e(n-n_e) \label{eq:ion}
\end{eqnarray}

\noindent where $n$ is the ion fluid density, $\beta_i$ is its
velocity, $m_i$ is the ion mass and $E_x$ is the charge separation
field. The density distribution of the electrons~($n_e$) is
calculated from the evolution of the CEL. We simply assume the
electrons from the front target are piled up, compressed and
uniformly distributed within a region with the size of the skin
length $l_s=1/\sqrt{n_0}$. We do this because the CEL is always in
front of the accelerated ions at this early stage. This assumption
will become invalid as soon as the ions catch up with the CEL. Our
calculation ends before this time and makes sure the hydrodynamic
velocity of the ions at the density peak point is larger than the
CEL velocity. The moving distance of the CEL at this time is just
the one we are looking for and it equals the minimum thickness of
the target for the ion acceleration~($l_{hydro}$). As we will see
usually this gives a smaller value than the usually used
one~($l_{immo}=a/{\pi}n$) obtained in the immobile ion model.

We solve the system of equations (\ref{CEL})-(\ref{eq:ion})
numerically. In our calculations, we vary the ion mass, target
density and laser intensity to see their effects on the minimum
target thickness. The foil is  assumed to be thick enough
initially~($0.2\lambda<x<1.8\lambda$) for the ions to catch up with
the CEL. The electron and ion density distributions at the
calculation end are shown in Fig.~\ref{pre-hole-boring}(a). The
density distributions also indicate one of the necessary criterions
to end our calculation: the time when the peak of the ion density
distribution reaches the CEL. After this the charge separation field
for CEL will deviate obviously from the one used in Eq.~(\ref{CEL}).
The moving distance of the CEL at this time is assumed to be the
minimum thickness of the target $l_{hydro}$ since the ions can catch
up with the CEL once the target is thicker than $l_{hydro}$.
Obviously we can get $l_{hydro}$ from the numerically calculations
show in Fig.~\ref{pre-hole-boring}(a), it can also be gotten by
considering the force balance as we will show in the following.
Fig.~\ref{pre-hole-boring}(b) shows the ratio of the CEL
velocity~($\beta_e$) at the end of the calculation to the
theoretical relativistic hole boring
velocity~$\beta_h=a/(a+\sqrt{m_in_0})$~\cite{Robinson2009}. As we
see the value tends to a constant depending only on the ion mass.
With this constant value it is then easy to calculate the
displacement of the CEL at this time from the relationship of the
almost balance between the charge separation force and the laser
pressure:

\begin{equation}\label{balance}
    {\pi}n_0^2l_1(l_1+l_s) = \frac{a_0^2}{{\pi}}\frac{1-\beta_e}{1+\beta_e}
\end{equation}

\noindent The force balance can be seen from
Fig.~\ref{pre-hole-boring}(c). The black line indicates the force
due to laser pressure, the red line indicates the charge separation
force and the blue line indicates the force due to mass increase.
The first two forces balance with each other very quickly once the
CEL moves $0.1\lambda$ into the target. The displacements of the CEL
obtained in Fig.~\ref{pre-hole-boring}(a) fit well with those gotten
from the force balance calculation. The latter is in
Fig.~\ref{pre-hole-boring}(d) with different ion masses. The usually
used value based on the immobile ions model~$l_{immo}$ is also shown
with the black line. As we see, when the intensity of the laser
electric field is larger than $a_0=100$, the present
results~$l_{hydro}$ are smaller than the value based on immobile
ions~$l_{immo}$. The lighter the ion mass, the larger the difference
with the one of the immobile ion model. $l_{hydro}$ is just the
minimum thickness of the foil target for ion acceleration. When the
target is thinner than this minimum value, ions can not catch up the electrons and
neutralize them. Then electrons are smashed away from the ions
completely by the light pressure and the target is transparent to
the laser later. The electrons are dispersed by the pulse and the
acceleration structure disappears. From the present calculation, we
see that ions have already caught up with the electrons before the
CEL moves a distance of $l_{immo}$ and the CEL will not completely
separate from the ions. So our model shows that the usual
value~$l_{immo}$ overestimates the minimum thickness. This finding
is especially important for the selection of the target thickness
for multicascade ion acceleration scheme recently proposed by
Gonoskov \textit{et al.} ~\cite{Gonoskov2009}.

\section{Ion acceleration by use of shaped foil target}\label{sec_shapetarget}

\begin{figure}
\includegraphics[width=6cm]{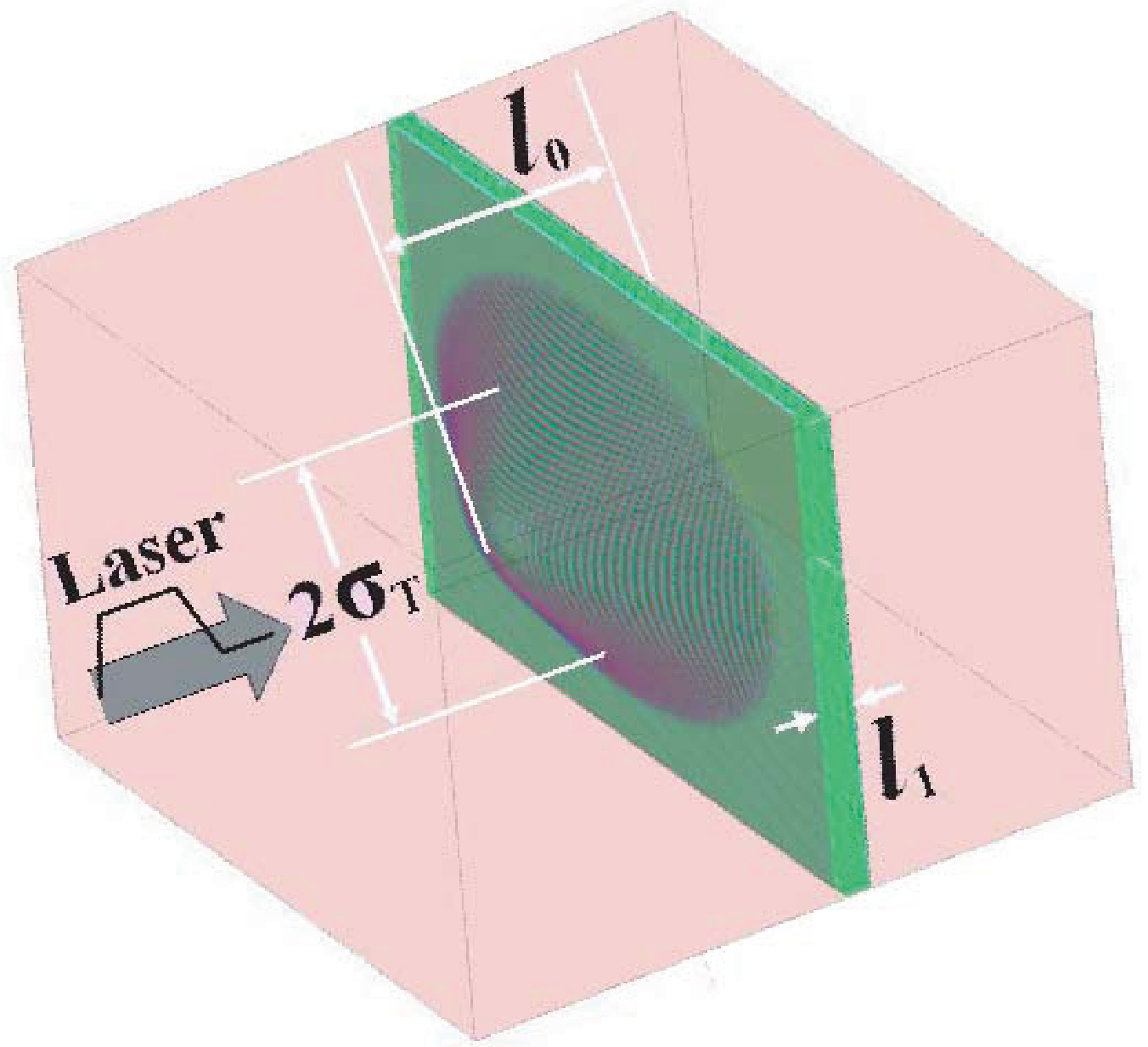}%
\caption{\label{layout} Layout of the interaction scheme. $\sigma_T$
defines the transverse Gaussian profile of the shaped target, $l_0$
is the maximal target thickness while $l_1$ is the cutoff thickness.
A circularly polarized laser pulse is incident on the foil target from
the left boundary.}
\end{figure}

Up to now we discussed the minimum target thickness for ion
acceleration in the laser pressure dominated regime. Once the target
is thicker than the minimum value, the whole target is opaque to the
laser pulse and it is accelerated in a hole-boring process with the
velocity $\beta_h$. When the CEL arrives at the rear of the foil,
the acceleration changes to a light sail process. Then, the momentum
evolution of the target satisfies~\cite{Robinson2008,Pegoraro2007}:

\begin{equation}\label{acce}
\frac{dp}{dt} =
\frac{2I}{c}\frac{\sqrt{p^2+\sigma^2c^2}-p}{\sqrt{p^2+\sigma^2c^2}+p},
\end{equation}

\noindent where $I$ is the laser intensity, $\sigma$ is the target area
density. For the velocity evolution of the target one obtains:

\begin{equation}\label{velocity}
\frac{d\beta}{dt} =
\frac{1}{2{\pi}n_0m_ic}\frac{E^2(t,x,r)}{l_0}\frac{1}{\gamma^3}\frac{1-\beta}{1+\beta}.
\end{equation}

\noindent Here $E$ is the laser electric filed amplitude, $n_0$ and
$l_0$ are the target initial density and thickness, respectively.

Eq.~(\eqref{velocity}) shows that the energy spread of accelerated
ions depends on the transverse variation of the local ratio of laser
intensity to the target area density. The distance the ions pass
under the laser pressure is $s(r){\propto}{E^2(t,x,r)}{l^{-1}_0}$.
An initially flat target is inevitably deformed, if the laser
intensity is not uniform transversely. The target deformation
quickly destroys the acceleration structure and deteriorates the
beam quality. From Eq.~(\ref{velocity}) we see that a target can be
uniformly accelerated if its areal density $\sigma$ is shaped
properly. For the usual transversely Gaussian
pulse~[$a=a_0{\times}exp(-r^2/\sigma_L^2)$], one can use a target
with the Gaussian thickness distribution as shown in
Fig.~\ref{layout}. In the following simulations, the distribution of
the target thickness along the transverse direction is:

\begin{equation}\label{thickness}
    l=max\{l_1,l_0{\times}\exp[(-r^2/\sigma^2_T)^m]\}.
\end{equation}

\noindent Here $r$ is the transverse distance to the laser axis,
$l_1,l_0,\sigma_T, m$ are the shape factors, which are marked in
Fig.~\ref{layout}.

\begin{figure}[t]
\includegraphics[width=9cm]{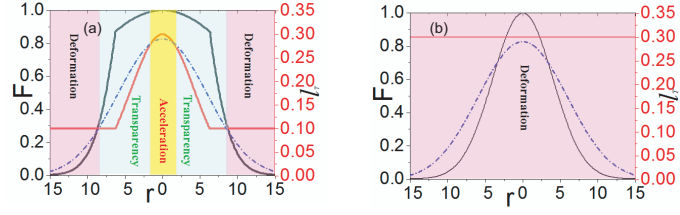}%
\caption{\label{division} Target partitions in the cases of (a) shaped
target and (b) flat target, according to the transparency
calculation. The parameters for the shaped target are:
$l_0=0.3\lambda, l_1=0.1\lambda, m=1, \sigma_T=6\lambda$. For the
flat target we just use $l_1=l_0=0.3\lambda$. The laser pulse has a
focus of $\sigma_L=8\lambda$. The red solid line represents the
target thickness distribution along the transverse direction; the
blue dashed line represents the minimum target thickness requirement
for an opaque target. The black line represents the acceleration
factor.}
\end{figure}

Before carrying out the simulations for such kind of shaped target,
we first check the target transparency for the pulse with a
transversely Gaussian pulse. In Fig.~\ref{division} we show the
minimum thickness requirement from theoretical calculation by use of
Eq.~(\ref{balance}) and the acceleration factor $F_{acc}=a^2/l$ in
the target transverse direction. $\beta_e$ is the function of
$a,n,m_i$ and we get the value from Fig.~\ref{pre-hole-boring}(b).
The shaped target thickness and flat target thickness distributions
are also shown in the figure. As we see from Fig.~\ref{division}(a)
for the shaped target, in the center it is thicker than the minimum
value~$l_{min}$. At this region, the acceleration factor $F_{acc}$
is also almost uniform, so the target can be accelerated uniformly.
Outside of this region, the target thickness is thinner than
$l_{min}$. The target is transparent to the laser pulse here, ions
can not get an effective uniform acceleration. In the outside
region, the target is thicker than $l_{min}$ again and $F_{acc}$
decreases with radius, so in this region the target is opaque and
will be accelerated and deformed. However, for the usual flat target
[see Fig.~\ref{division}(b)], the target thickness is always larger
than the minimum value, so it is opaque to the laser pulse and the
whole target belongs to the deformation region. We will demonstrate
these in the following PIC simulations.

By use of Eq.~(\ref{balance}) we can also get the final size of the
accelerated ion bunch in the target center . For the present target
shape the calculated bunch radius is:
$r_b=\sigma_L\sigma_T\sqrt{\ln[{\pi}n_0l_0a_0^{-1}\sqrt{(1+\beta_e)/(1-\beta_e)}](\sigma_L^2-\sigma_T^2)^{-1}}$,
where $\beta_e=\alpha\beta_h$ and $\alpha$ is a function of ion mass
and laser intensity. Theoretically it can be obtained from
Fig.~\ref{pre-hole-boring}(b) in which $\alpha\approx0.5$. However
in the PIC simulations we find the accelerated bunch size can be
approximated much better when we choose $\alpha\approx 1$. This
difference may result from our too strict criterion for the minimum
target thickness. From the calculated bunch size we can also get
requirements for the shape factors, such as $\sigma_T<\sigma_L$ and
$l_1{\leq}l_0{\times}\exp(-r_b^2/\sigma_T^2)$. In the next section
we present results of multi-dimensional PIC simulations, compare
them with the above results and use them to get the optimal shape
factors.

\subsection{Particle-in-cell simulations}

We use the VLPL code to do both 2D and 3D
simulations~\cite{Pukhov-vlpl}. First, we do 3D-simulation to show
the ion acceleration from the shaped foil target. and then we
investigate in detail the effects of the target parameters on the
beam quality by less time-consuming 2D simulations.

\begin{figure}
\includegraphics[width=9cm]{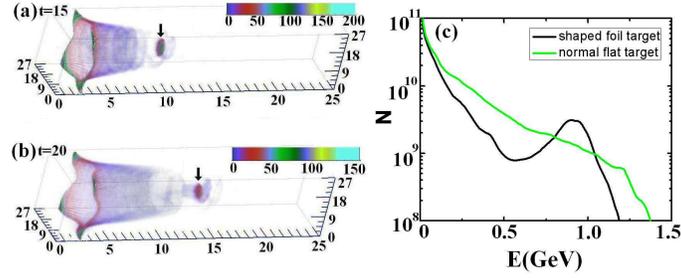}%
\caption{\label{3dsim}Proton density distributions in three
dimensional simulations at $t=15T_0$ (a) and $20T_0$ (b). The
single-headed arrow indicates the high quality proton bunch with a
higher energy and better collimation. The initial target with
$\sigma_T=6\lambda$ is located between $x=2.0\lambda$ and
$2.3\lambda$, irradiated by a CP laser pulse with
$\sigma_L=8\lambda$. (c) Proton energy spectra at $t=20T_0$. Here,
the normal flat target with $l_0=l_1=0.3\lambda$ is located at the
same position as the shaped foil target and is irradiated by the
same laser pulse.}
\end{figure}

The simulation box in the 3D-simulation is $25\lambda \times27\lambda
\times27\lambda$, which consists of $2500 \times 225 \times225$
cells. A circularly polarized laser pulse with a Gaussian profile in
space and a trapezoidal profile (linear growth - plateau - linear
decrease) in time is normally incident on the foil target. The
temporal profile satisfies:

\begin{equation} \label{eq:4}
a=\left\{ \begin{aligned}
         & a_0\exp(-\frac{y^2}{\sigma_L^2})t, \quad 0\leq t<1T\\
         & a_0\exp(-\frac{y^2}{\sigma_L^2}),\quad 1T\leq t\leq6T,\\
         & a_0\exp(-\frac{y^2}{\sigma_L^2})(8-t),\quad 6T<t\leq7T,
         \end{aligned} \right.
\end{equation}
\noindent where $a_0=100$ is the normalized intensity of the laser
electric field, $\sigma_L=8\lambda$ is the focal spot radius,
$T=3.3fs$ is the laser cycle for an infrared laser with $1~\mu$m
wavelength. The initial target is located between $x=2.0\lambda$ and
$2.3\lambda$, with a transversely varying thickness, as shown in
Fig.~\ref{layout}. Here, the cutoff thickness $l_1=0.15\lambda$ and
$\sigma_T=6.0\lambda$. The plasma density is $n_0=100$.

Fig.~\ref{3dsim} presents the proton density distributions at two
time points: $t=15T_0$ and $20T_0$. One sees that the center part of
the target is accelerated strongly and soon breaks away from the
whole target. As expected, the deformation of the target center is
well suppressed and a peak appears in the proton energy spectrum,
which is shown in Fig.~\ref{3dsim}(c). The number of the protons
whose energy is larger than 0.65GeV is about $8.3\times10^{11}$ and
their total energy is about 120J. Instead, for a usual flat target,
the energy spectrum shows a typical exponential decay due to the
easy heating and deformation of the target. By diagnosing the
divergency angle distribution, the average divergency of these
energetic protons from the shaped foil target is less than
$5^\circ$. The whole target can be stably accelerated until the end
of laser irradiation with the energy conversion efficiency as high
as $23.1\%$.

In Fig.~\ref{3dsim}(a,b), it is easy to distinguish the
deformation region, acceleration region and transparency region. The
center part is the acceleration region, which composes the energy
peak in the spectrum. Around it is the transparency region, where the
electron density is quickly low enough so that the laser pulse can
easily penetrate it and then go through it. This region separates
the acceleration region from the deformation region locating in the
outer side of the target and effectively suppresses the target
heating and protects the acceleration region. For a
flat target, this transparency region can also be formed due to the
density dilution during the target deformation. However, the process
is much slower and
it is a gradual changing. Target heating is inevitable then,
and there is no obvious acceleration region formation. The
pre-shaped target makes the final isolated acceleration bunch
possible. The radius of the bunch in our simulation is around
$4.5\lambda$, which is close to the theoretical estimate for $r_b$.

Although our 3D simulations show the possibility of a GeV
monoenergetic proton beam acceleration, a well shaped
target is required. In experiment, it may be difficult to make a well matched target
without any deviations. Certainly, the final beam quality should be
related with the target shape factors. For real applications, we
are going to demonstrate the robustness of this shaped target scheme.
In the following we discuss in detail the effects of these factors on the
final beam bunch. To save the computational time, we only perform 2D
simulations to show the effects. A series of 2D simulations have
been performed. The whole simulation box is
$32\lambda\times32\lambda$, sampled by $3200\times3200$ cells. The
foil target initially locates between $x=5.0\lambda$ and
$5.3\lambda$. The CP laser pulse has the same profile both in time
and space as above 3D case except the pulse duration is now
$\tau=10$, which corresponds to a distribution of 1T-8T-1T in time.

\subsection{Dependence on the shape factor}
\begin{figure}
\includegraphics[width=9cm]{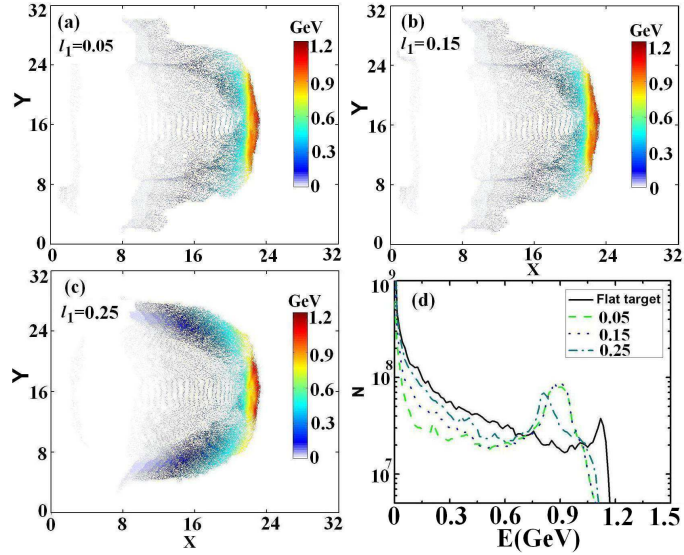}%
\caption{\label{cutoff} Proton energy distributions in x-y space for
different cutoff thickness $l_1$: (a) $0.05\lambda$, (b)
$0.15\lambda$ and (c) $0.25\lambda$. All the other laser pulse and
target parameters are the same. (d) Proton energy spectra at
$t=30T_0$. Here the flat target refers to the one with $n_0=100$ and
$l_0=l_1=0.3\lambda$.}
\end{figure}

We first take into account the influence of the cutoff thickness
$l_1$ on the beam quality. In the simulation we fix all other
parameters and only change $l_1$. The ratio of the target width to
the laser focus~($\sigma_T/\sigma_L$) is kept to be 7/8.
Fig.~\ref{cutoff} shows the simulation results. We can see that the
spatial energy distribution in Fig.~\ref{cutoff}(a) and (b) are
almost the same. The cutoff thicknesses for these two cases are
$0.05\lambda$ and $0.15\lambda$, respectively. The corresponding
energy spectra are shown in Fig.~\ref{cutoff}(d). Again, the energy
distributions of high energy protons for these two cases are the
same. However, when $l_1$ is increased to $0.25\lambda$, the energy
spectrum changes significantly as shown in Fig.~\ref{cutoff}(d). The
peak energy decreases and the cutoff energy increases. The spectrum
tends to be the one of a flat target. It shows there exists a
threshold value for $l_1$. When $l_1$ is larger than the threshold,
the spectra are significantly different. In additional simulations, we find
that the threshold is about $0.20\lambda$ in the present case. This
is close to the theoretical value of our analysis above. When the cutoff thickness is
smaller than the $l_0\times{\exp}(-r_b^2/\sigma_T^2)$, the
accelerated bunch size is almost constant as shown in
Fig.~\ref{cutoff}(a) and (b). When it increases, no obvious
transparency region separates the acceleration and the deformation
regions. Target deformation  happens continuously along the target
and the effectively accelerated bunch is smaller as shown in
Fig.~\ref{cutoff}(c) and the final spectrum is close to the flat
target case.

\begin{figure*}
\includegraphics[width=14cm]{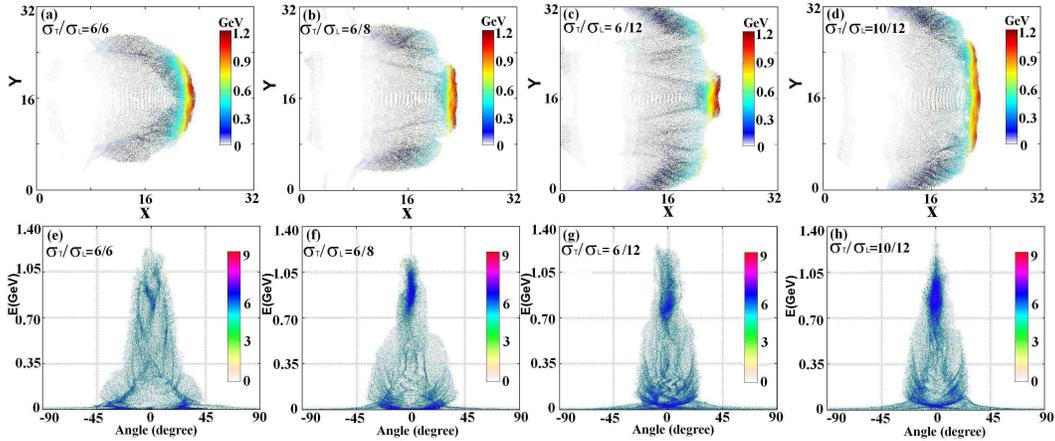}
\caption{\label{2dsigma} Proton energy distributions in x-y space
for different $\sigma_L$ and $\sigma_T$ at $t=30T_0$: (a)
$\sigma_L=6\lambda$, $\sigma_T=6\lambda$, (b) $\sigma_L=8\lambda$,
$\sigma_T=6\lambda$, (c) $\sigma_L=12\lambda$, $\sigma_T=6\lambda$,
and (d) $\sigma_L=12\lambda$, $\sigma_T=10\lambda$. Corresponding
proton energy distributions as a function of the divergency angle
are shown in (e)-(h).}
\end{figure*}

The most important factors are the matching parameters: $\sigma_T$
and $\sigma_L$. In the following we check their effects on the beam
quality. Fig.~\ref{2dsigma} shows some typical simulation results
for different $\sigma_T$ and $\sigma_L$. The top four figures show
the proton energy distributions in space while the bottom four
correspond to the angular distributions. We fix $\sigma_T=6\lambda$
and increase $\sigma_L$ from $6\lambda$ to $12\lambda$. It is shown
that, when $\sigma_T$ is close to $\sigma_L$, the target
deformation happens. Most protons are located at the deformation
region and the target evolves into a natural cone. The corresponding
energy-divergency distribution is widely spread, as shown in
Fig.~\ref{2dsigma}(e). With the increase of laser focus, the center
part of the target is uniformly accelerated so that it can break
away from the whole target. The three regions mentioned before can
be easily distinguished from the Fig.~\ref{2dsigma}(b-c). A bunch of
protons with a higher energy and a better collimation is formed in
Fig.~\ref{2dsigma}(f-g). The radius of the bunch decreases with
$\sigma_L$, which confirms the theoretical analysis. When
$\sigma_L$ increases, the transparency region extends to the target
center, a larger laser focus (or laser energy) leads to a smaller
accelerated bunch. The results show the importance of a well matched
target. Fig.~\ref{2dsigma}(d) and (h) correspond to a well matched
case when the laser focus is $\sigma_L=12\lambda$. When we increase
the target width close to the $\sigma_L$, the acceleration region
broadens and more ions are uniformly accelerated.
\begin{figure}[b]
\includegraphics[width=9cm]{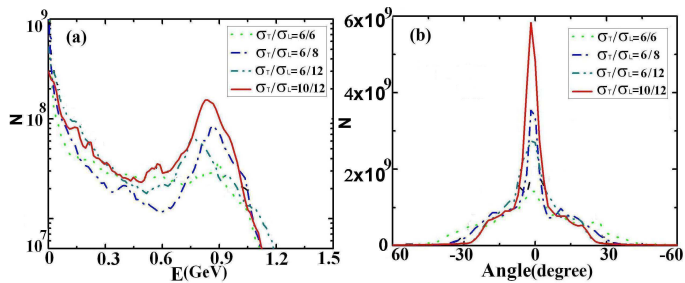}%
\caption{\label{2dspec} Proton energy spectra (a) and divergency
angle (b) distributions for different laser focus radii $\sigma_L$ and
$\sigma_T$ at $t=30T_0$.}
\end{figure}
Fig.~\ref{2dspec} shows both the energy spectra and the angular
distributions for these cases. As expected, there is a clear
quasi-monoenergetic peak both in the case with
$\sigma_T/\sigma_L=6/8$ and $\sigma_T/\sigma_L=10/12$. The peak
energy is about 0.85GeV and 0.80GeV, respectively. The corresponding
full-width of half maximum divergency angle is about $6^\circ$ and
$4^\circ$. Obviously, for the well matched cases the larger the
laser focus, the more the accelerated protons. On the contrary, for
the not perfect matched case, both the peak energy and the total
production of the accelerated protons decrease. For the unmatched
case, no clear peak appears and the proton number decreases further.
\begin{table}
\caption{Available and optimum values of $\sigma_T/\sigma_L$}
\label{tab-sigma}       
\begin{tabular}{|c|cccccc|c|}
\hline
&\multicolumn{7}{c}{$\sigma_T/\sigma_L$}\vline \\
\cline{2-8}
 \raisebox{2.3ex}[0pt]{$\sigma_L$}&\multicolumn{6}{c}{Available values}\vline&Optimum \\
\hline
6 & 0.5 & 0.583 & 0.677 & 0.75 & 0.833 & 0.916 & 0.75\\
\hline
8 & 0.375 & 0.5 & 0.6 & 0.75 & 0.8125 & 0.875 & 0.8125\\
\hline
10 & 0.4 & 0.5 & 0.6 & 0.7 & 0.8 & 0.9 & 0.8\\
\hline
\end{tabular}
\end{table}

In order to obtain the optimal ratio of $\sigma_T/\sigma_L$ in the
simulation, we perform the parameter scanning as shown in
Table~\ref{tab-sigma}. Here, all the target and laser parameters are
the same except for $\sigma_T$ and $\sigma_L$. The available values of
$\sigma_T/\sigma_L$ mean that a high quality proton bunch with a
quasi-monoenergetic peak and low divergency angle can be observed
with these parameters. The optimum value indicates the best bunch
quality such as the narrowest energy spread and the lowest divergency. It
is shown that the tolerable values of $\sigma_T/\sigma_L$ exist
around 0.50-0.90 while the optimum value is about 0.80. These
simulations supplement our analytical results which only give the
condition of $\sigma_T/\sigma_L<1$ and also give some quantitative
illumination to the experiments.

\subsection{Effect of target surface roughness}

Since in our scheme, the target thickness is less than the laser
wavelength, i.e. nanometer-thickness, the relatively larger surface
roughness of the target might be inevitable in real experiments
and it may have influence on the final accelerated ion beam. Here we
check its effects by comparing three simulations with different
surface roughness: (a) a smooth surface, (b) $10\%$ roughness and
(c) $30\%$ roughness. The initial targets are shown in
Fig.~\ref{2droughness}(a). In order to resolve the surface
roughness, both the longitudinal and transversal cell sizes should
be small enough which leads to extremely small steps both in
space and time in the simulation. This makes the simulations extremely
time consuming.
Therefore, we only present the simulation results at an early time
$t=10T_0$. This time is, however, already long enough to see the final
effects.
Fig.~\ref{2droughness}(c) shows proton energy spectra for
these cases. We notice that all the spectra show a clear energy peak
despite the different surface roughness. Yet, for the
target with 30\% surface roughness, the peak energy is about
0.25GeV, which is higher than the value of 0.2GeV in the cases
with a lower roughness. Similarly, the cutoff energy is also higher
than the other two cases. The differences between the two lower roughness
cases are much smaller. The main effect of the target roughness is
increasing the laser absorption and conversion efficiency of its
energy to the super hot electrons. These electrons are easily
dispersed in space and awake the TNSA acceleration. This can be seen
in Fig.~\ref{2droughness}(b), in which the energy spectrum of the
electrons is shown. Obviously the target with $30\%$ roughness has a
much higher electron temperature. The other two cases are similar.
In addition to the spectrum we also check the angular distribution. The results
are shown in Fig.~{\ref{2droughness}(d). There is no obvious
difference with the case of smooth target and a target with $10\%$
roughness. So generally speaking, the simulation shows the roughness
of $10\%$ is acceptable. It gives a quantity guidance for
experimental demonstrations.

\begin{figure}
\includegraphics[width=9cm]{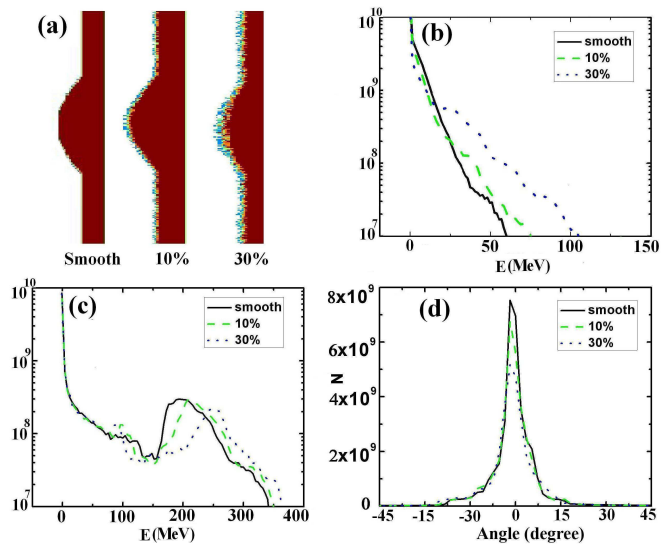}%
\caption{\label{2droughness} Shaped foil targets with different
surface roughness (a). Electron energy spectra (b) and proton energy
spectra (c) for different surface roughness at $t=20T_0$. Proton
divergency angle distributions at $t=20T_0$ (d). Here, the cutoff
thickness for these cases is $0.15\lambda$.}
\end{figure}

\subsection{Radiation damping effect}
Further, we perform simulations to check the radiation damping
effects, which was found to be very important for transparent nano
targets~\cite{Kulagin2007}. To simulate the damping effect we
suppose that, at any given moment of time, the electron radiation
spectrum is synchrotron like~\cite{kiselev-synchrotron}. The
critical frequency $\omega_c$ is given by the relation
$\omega_c=(3/2)\gamma^2|{\bigtriangleup}P_\perp|/(mcdt)$;
${\bigtriangleup}P_\perp$ is the variation of transversal electron
momentum force during the time step of $dt$. In our PIC code, we
follow trajectories of each electron and calculate the emission
during the interaction. We calculate the damping effects by
considering the electron's recoil due to the emitted radiation. The
recoil force was included in the equations of electron motion. We do
3D simulations with the same parameters as mentioned above,
while the radiation module switched on this time.

Fig.~\ref{3ddamping} shows the synchrotron radiation distribution.
$\theta$ is the angle between the directions of photon emission and
laser propagation; $\phi$ is the azimuthal angle. We find that the
emitted x-ray photons are mainly radiated at the angle of
($\theta=35^o-45^o$). Totally 8J energy~(about $1.54\%$ of the laser
energy)~is transferred to the x-ray photons with the average photon
energy about 1.2MeV. For the electrons' cooling, our simulation
shows that there is no obvious changing in the final electron energy
spectrum. Only the highest electrons~(400MeV$<E_k<$800MeV)~are a
little cooling down. The total electron energy is about 0.123J less
than the one in the simulation without the radiation effect. This
means the electrons can quickly get back their radiation-lost energy
from the driving pulse. For the ion acceleration, our simulation
results show there is almost no observable difference in the final
ion spectra between the cases with and without radiation damping
effects. Obviously this little electron cooling can not benefit the
ion acceleration so much for these laser-plasma parameters.

\begin{figure}
\includegraphics[width=7cm]{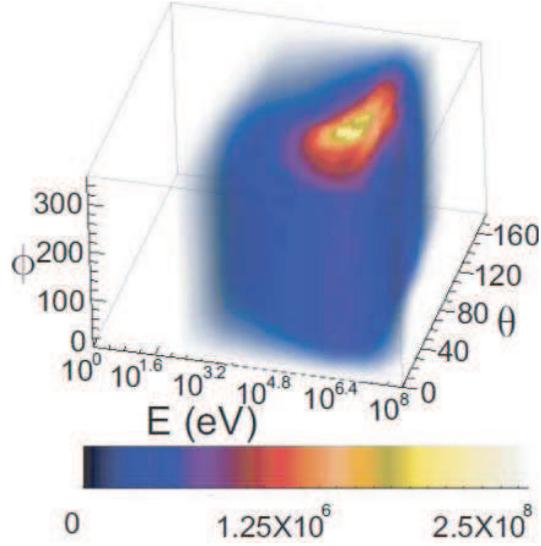}%
\caption{\label{3ddamping} Synchrotron radiation distribution in the
angular-energy space at $t=20T_0$. The colorbar represents the
photon number.}
\end{figure}

\section{Discussion and Summary}

We should point out that the target shaping only helps
to reduce electron heating and keeps the acceleration much more
uniform and for a longer time. However, the instabilities still
exist~\cite{Pegoraro2007,Chen2008} in the accelerated plasmas.
Although under perfect matching condition proton numbers can be
increased by increasing the laser focus as Fig.~\ref{2dsigma}(d)
suggests, the surface instability will develop after some time and destroy
the acceleration structure, which limits the final proton energy.
Suppression of such kinds of instabilities should be an important
work both for the laser ion acceleration itself and for the fast
ignition of inertial fusion targets based on laser-accelerated ion
beams~\cite{fusion}. These are important issues for the future work.

{In summary, by multi dimensional PIC simulations, we have checked
the target shape effects on GeV mono-energetic proton beam
acceleration. We propose to use a shaped target to match the laser
intensity. By this, there will be a transparency region separating
the acceleration region and deformation region and keeping the
acceleration structure for a longer time compared with a usual flat
target. Final spectrum shows a well mono-energetic peak once the
well shaped target is used. Effects of shape factors and target
surface roughness are also checked, which demonstrates the
robustness of our scheme. Finally, synchrotron radiation and damping
are checked. It does little influence on the ion acceleration for
the present parameter region.}

\begin{acknowledgments}\suppressfloats
This work is supported by the DFG programs TR18 and GRK1203. MC
acknowledges support by the Alexander von Humboldt Foundation. TPY
thanks the scholarship awarded by China Scholarship Council(CSC NO.
2008611025). ZMS is supported in part by the National Nature Science
Foundation of China (Grants No. 10674175, 60621063) and the National
Basic Research Program of China (Grant No. 2007CB815100).
\end{acknowledgments}


\end{document}